\documentclass[%
 reprint,
 amsmath,amssymb,
 aps,
]{revtex4-2}

\usepackage[english]{babel}         
\usepackage[utf8]{inputenc}         
\usepackage[T1]{fontenc}            
\usepackage{lmodern}                
\usepackage{textcomp}               
\usepackage{microtype}              
\usepackage[top=3cm, bottom=3cm, inner=3cm, outer=3cm]{geometry}
\usepackage{setspace}
\usepackage{parskip}
\usepackage{amsmath}                
\usepackage{amssymb}                
\usepackage{mathtools}              
\usepackage{commath}                
\usepackage{dsfont}                 
\usepackage{bm}                     
\usepackage{siunitx}
\usepackage{cancel}
\usepackage{braket}
\usepackage{diffcoeff}
\usepackage[version=4]{mhchem}

\usepackage{graphicx}               
\usepackage{float}                  

\usepackage{diagbox}                
\usepackage{booktabs}               

\usepackage{csquotes}

\usepackage{enumitem}               
\usepackage{xcolor}                 

\usepackage[yyyymmdd]{datetime}     

\usepackage[
    hidelinks,
    linktoc     = all,
    colorlinks  = false,
    filecolor   = blue,
    linkcolor   = blue,
    urlcolor    = blue,
    citecolor   = blue,
    anchorcolor = blue,
]{hyperref}

\usepackage{glossaries-extra} 
\glsdisablehyper

\usepackage{bookmark}                                   
\usepackage{url}                                        
\usepackage[capitalize, noabbrev, nameinlink]{cleveref} 

\crefname{equation}{}{}  

\renewcommand{\Re}{\textnormal{Re}}

\difdef{f,s,c}{}{
  op-symbol = \mathrm{d},
}
\difdef{f,s}{f}{
	op-symbol = \delta,
}
\difdef { l } { dn } { style = d^ }

\newcommand{\fom}{f_\text{FOM}} 
\newcommand{\dvf}{p}            
\newcommand{\fop}{(\vec r)}     
\newcommand{\fopp}{(\vec r')}   
\newcommand{\foppp}{(\vec r'')} 
\newcommand{\iop}{\dl.dn.[3]{\vec r}}     
\newcommand{\iopp}{\dl.dn.[3]{\vec r'}}   
\newcommand{\ioppp}{\dl.dn.[3]{\vec r''}} 
\renewcommand{\vec}{\mathbf}

\newcommand{\gom}{\ensuremath{g_\text{OM}}}
\newcommand{\omscat}{\ensuremath{\Gamma_\text{OM}}}

\newcounter{subfigure}[figure]

\newcommand{\sublabel}[1]{%
    \refstepcounter{subfigure} \label{#1}
}

\newabbreviation{xhope}{X-HOPE}{X-point matching by harmonic period extension}
\newabbreviation{omc}{OMC}{optomechanical crystal}
\newabbreviation{vna}{VNA}{vector network analyzer}
\newabbreviation{sa}{SA}{spectrum analyzer}
\newabbreviation{eom}{EOM}{electro-optic modulator}
\newabbreviation{fom}{FOM}{figure-of-merit}
\newabbreviation{pd}{PD}{photodetector}
\newabbreviation{voa}{VOA}{variable optical attenuator}

\usepackage[tickmarkheight=.5em,textwidth=\marginparwidth,textsize=footnotesize]{todonotes}

\begin{document}

\preprint{APS/123-QED}

\title{Inverse-designed release-free optomechanical crystal \\ with high photon--phonon coupling}

\author{David Hambraeus}%
 \email{davham@chalmers.se}%
\author{Paul Burger}%
\author{Johan Kolvik}%
\author{Philippe Tassin}
\altaffiliation{
 Department of Physics, Chalmers University of Technology. 41298 Göteborg, Sweden}%
\author{Raphaël Van Laer}%
\email{raphael.van.laer@chalmers.se}%
\affiliation{%
 Department of Microtechnology and Nanoscience (MC2), Chalmers University of Technology. 41298 Göteborg, Sweden}%

\date{\today}

\begin{abstract}
    Interaction between light and mechanics provide a powerful interface between optical and microwave signals, with applications spanning classical signal processing and quantum technologies. High-performance optomechanical devices require both strong photon--phonon coupling and tolerance to parasitic laser heating. Release-free optomechanical crystals provide improved thermal anchoring compared to suspended nanobeams, but have so far exhibited weaker vacuum optomechanical coupling rates, leaving a trade-off between coupling strength and thermal robustness. Here, we largely close this gap: we design and experimentally demonstrate a release-free silicon optomechanical crystal with a record vacuum optomechanical coupling rate of about $\gom / (2 \pi) = \qty{800}{\kilo\hertz}$, comparable to suspended state-of-the-art devices. The resulting optomechanical scattering rate $\omscat/(2 \pi)=\qty{1.1}{\kilo\hertz}$ is nearly twice that of previous release-free implementations. This performance is achieved by combining physics-guided human intuition with a multiphysics inverse-design algorithm introduced here for resonant optomechanical structures. Beyond the specific device demonstrated, the inverse-design framework is applicable to co-optimizing optical and mechanical resonances and eigenmodes more broadly. These results strengthen release-free optomechanical crystals as a platform for fast, low-noise classical and quantum optomechanics.
\end{abstract}

\maketitle

\section{Introduction}

Optomechanical interaction in solid-state systems enable strong coupling between optical fields and mechanical motion \cite{aspelmeyer_cavity_2014}, with applications ranging from precision sensing \cite{schliesser_high-sensitivity_2008, sansa_optomechanical_2020, gavartin_hybrid_2012} to microwave–optical transduction \cite{mirhosseini_superconducting_2020, jiang_optically_2023} and quantum information processing \cite{shandilya_optomechanical_2021}. These interactions have enabled key demonstrations such as slow light \cite{safavi-naeini_electromagnetically_2011}, laser cooling of mechanical resonators to their ground state \cite{chan_laser_2011}, and the generation of squeezed states of light and motion \cite{safavi-naeini_squeezed_2013}. High-performance optomechanical devices require both large photon–phonon coupling rates and robustness against optical absorption–induced thermal noise.

\Glspl{omc} achieve strong coupling by co-localizing optical and mechanical modes within wavelength-scale nanostructures \cite{chan_optimized_2012, sonar_high-efficiency_2025, safavi-naeini_controlling_2019}. To suppress mechanical radiation loss, these devices are typically implemented as suspended nanobeams. While suspension enables tight confinement, it also limits thermal anchoring, rendering suspended \glspl{omc} susceptible to excess noise under optical pumping \cite{meenehan_pulsed_2015, burger_design_2025, sonar_high-efficiency_2025}. Several approaches have been explored to mitigate this limitation, including two-dimensional suspended geometries \cite{ren_two-dimensional_2020,mayor_high_2025,sonar_high-efficiency_2025,safavi-naeini_two-dimensional_2014} and active cooling using superfluid helium \cite{qiu_laser_2020}.

Release-free optomechanical crystals provide an alternative architecture in which the device remains mechanically attached to the substrate, allowing heat generated by optical absorption to dissipate directly into the substrate \cite{kolvik_clamped_2023}. This geometry substantially improves tolerance to optical heating \cite{kolvik_clamped_2023,kolvik_optomechanical_2025}. However, first-generation release-free \glspl{omc} exhibited vacuum optomechanical coupling rates of approximately $\gom / (2 \pi) = \qty{500}{\kilo\hertz}$, leading to optomechanical scattering rates $\omscat = 4 \gom^2 / \kappa$, where $\kappa$ is the optical linewidth, below those of state-of-the-art suspended devices \cite{chan_optimized_2012, ren_two-dimensional_2020, mayor_high_2025, sonar_high-efficiency_2025}. This limitation arose because release-free operation requires a high-wavevector mechanical mode and an optical mode near the half-X-point rather than at a band edge \cite{kolvik_clamped_2023, burger_design_2025} -- a configuration in which the optical field extends gradually into the mirror, producing sub-optimal spatial overlap with the tightly confined mechanical mode.

In this work, we close this gap and demonstrate a release-free silicon optomechanical crystal with a vacuum optomechanical coupling rate of $\gom / (2 \pi) = \qty{800}{\kilo\hertz}$, comparable to state-of-the-art suspended devices. The device is realized through two complementary contributions: a physically motivated mode-engineering technique (\glsxtrlong{xhope}, or \glsxtrshort{xhope}\glsunset{xhope}) that restores spatial overlap between the optical and mechanical modes, and a multiphysics inverse-design algorithm for resonant optomechanical structures that recovers the radiation-limited quality factors \gls{xhope} alone leaves too low. Combined with the intrinsic light-resilience of the release-free architecture \cite{kolvik_optomechanical_2025}, these results establish release-free \glspl{omc} as a platform for fast, low-noise classical and quantum optomechanics, and provide a key ingredient for release-free piezo-optomechanical microwave-to-optical transducers \cite{burger_design_2025}.

\begin{figure*}[t]
    \centering
    \includegraphics[width=1.0\textwidth]{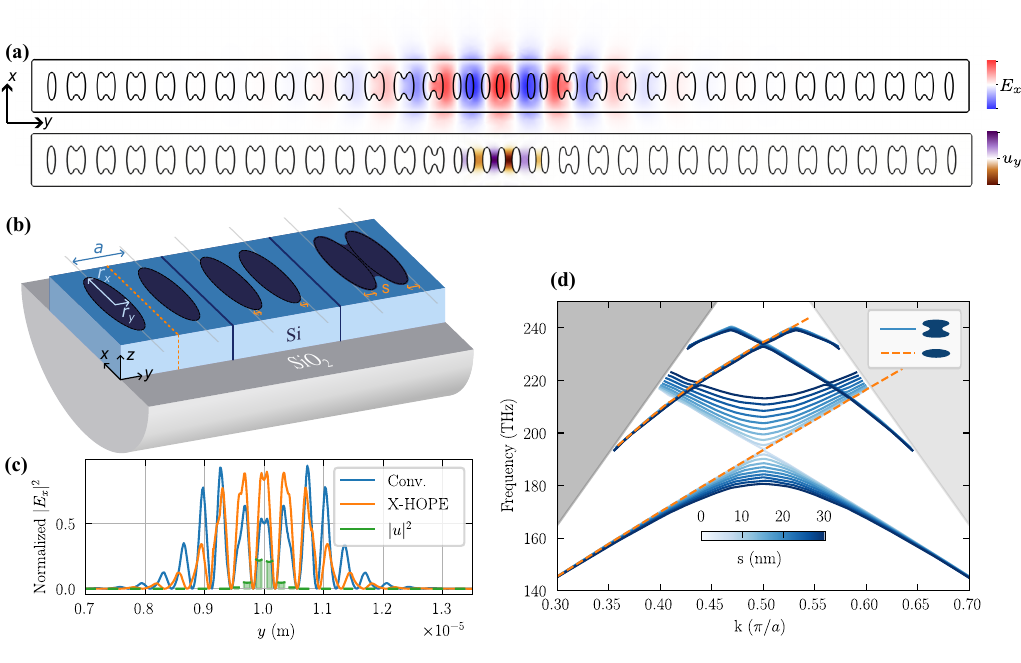}
    \caption{%
        \textbf{Release-free optomechanical crystal cavity with improved photon-phonon overlap.} 
        \textbf{(a)}
        The release-free \glsxtrshort{omc} with \glsxtrshort{xhope} design pattern showing
        the transverse component of the normalized electric field $E_x$ of the optical mode (top) and
        the longitudinal normalized displacement $u_y$ of the mechanical mode (bottom).
        \textbf{(b)}
        Diagram of unit cell geometry. The basic cell is an elliptical hole with major and minor radii $r_x$ and $r_y$. The
        double cell used for the mirror consists of two holes that have been moved
        toward each other by a distance $s$. When $s$ is large, the holes are merged to
        prevent bridges too thin to fabricate.
        \textbf{(c)}
        Optical intensity along the center of the nanobeam for a
        conventional mirror transition \cite{kolvik_clamped_2023} and our
        design. The displacement of the \glsxtrshort{xhope} mechanical mode is shown in
        green with arbitrary units (the mechanical mode shape is very similar for the conventional
        case). The \glsxtrshort{xhope} design has an
        increased optical intensity in the center of the beam, resulting in
        improved photon-phonon overlap.
        \textbf{(d)}
        Unit cell band structure for the double cell (shades of blue) and single
        cell (dashed orange). The x-axis is showing the $k$ in units of $\pi/a$,
        which would be the X-point of the single cell beam with periodicity $a$.
        The lighter (right) gray light-cone only appears when considering the double cell.
        At $s=0$, meaning no deviation from regularly spaced holes, the bands of
        the double cell are the combination of the single cell bands and a
        folded copy, effectively mirrored about $k=0.5\pi/a$. For $s>0$, a
        quasi-bandgap opens at $k=0.5\pi/a$, the new X-point.
    }
    \label{fig:x-hope_design}

    \sublabel{fig:fields}           
    \sublabel{fig:uc_params}        
    \sublabel{fig:conv_compare}     
    \sublabel{fig:uc_bands}         
\end{figure*}

\begin{figure*}[!t]
    \centering
    \includegraphics[width=1.0\textwidth]{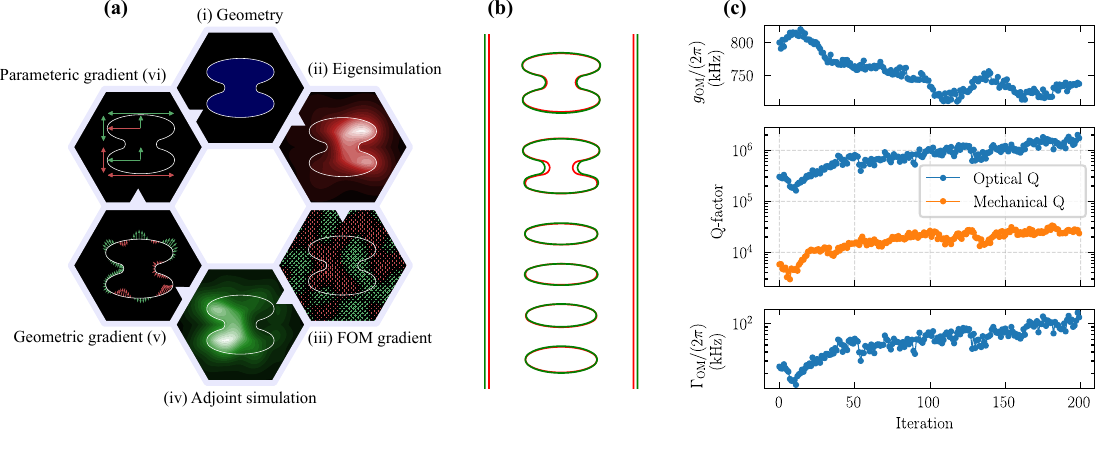}
    \caption{
        \textbf{Inverse-design of the optomechanical cavity enhances the optomechanical scattering rate.}
        \textbf{(a)}
        A diagram depicting the inverse design algorithm.
        (i) The geometry is imported into the simulation software.
        (ii) An eigenmode simulation is performed, yielding the eigenmode fields and frequencies.
        (iii) The gradient of the \glsxtrlong{fom} w.r.t.\ the fields is computed.
        (iv) An adjoint simulation is performed, yielding the adjoint fields.
        (v) Combining the adjoint and eigenmode fields, a geometric gradient is obtained.
        (vi) From the geometric gradient, the gradient w.r.t.~the design parameters is computed,
        which in turn is used to obtain an updated set of parameters and the cycle repeats until
        performance is satisfactory.
        \textbf{(b)}
        Comparison of the defect and transition region before (red) and after (green) optimization.
        \textbf{(c)}
        Evolution of simulated optomechanical interaction strength, optical and
        mechanical quality factors, and the optomechanical scattering rate $\omscat$ during optimization.
    }
    \label{fig:inverse-design}
    \sublabel{fig:id_concept}           
    \sublabel{fig:geom_change}          
    \sublabel{fig:optimization_trace}   
\end{figure*}

\section{Results}
\subsection{Design}
\label{sec:design}

\subsubsection{X-HOPE design technique}

Optomechanical crystal cavities enable strong light--sound interactions by confining both optical and acoustic fields to a wavelength-scale region. The central design challenge is to realize mirror regions, where a bandgap prevents the modes' propagation, and defect regions, where the modes can interact, simultaneously for near-infrared light and gigahertz sound while maintaining strong spatial overlap between the optical and mechanical modes. In addition, release-free \glspl{omc} rely on mechanical modes with large wavevectors to suppress leakage into the substrate and enable confinement without suspension, which increases the design complexity \cite{kolvik_clamped_2023,burger_design_2025}.

Strong optomechanical interactions further require phase-matching between the optical and mechanical waves. In existing release-free \glspl{omc}, this constraint prohibits the use of an optical mode at the X-point \cite{kolvik_clamped_2023,burger_design_2025}. Instead, the release-free \gls{omc} can be designed with an optical mode with wavevector near the half-X-point ($k = 0.5 \pi/a$, where $a$ is the periodicity) and a mechanical mode at the X-point (wavevector $k = \pi/a$), enabling phase-matched counter-propagating optomechanical interactions \cite{kolvik_clamped_2023,burger_design_2025}. As a result, the X-point bandgap for the optical mode is almost \qty{100}{\tera\hertz} from the cavity resonance, requiring a large change in geometry to shift the bandgap onto the optical mode frequency. During the initial phase of this transition, the optical field is not diminished and the beam only begins to act like a mirror once the bandgap fully overlaps the mode frequency. Consequently, the optical mode remains relatively large compared to the tightly confined mechanical mode, which suppresses the optomechanical interaction rate $\gom$.

Here, we address this challenge by extending the periodicity of the crystal to be twice as large by pairwise perturbation of the unit cells. Thus the X-point is abruptly moved to match the wavevector of the optical mode, even with an infinitesimal perturbation of the geometry. This could be accomplished by almost any perturbation that breaks the periodicity applied pairwise to the unit cells. In our structure, we move the holes closer together by a distance $s$ in a pairwise fashion (\cref{fig:uc_params}). This opens a bandgap where there was previously a continuous band, which confines the mode along the beam (\cref{fig:uc_bands}). To prevent too thin features, neighboring holes are merged when $s$ becomes large. While it is not necessary to make the perturbation so large that the merging is needed, a smaller $s$ (and thus a smaller bandgap) lets the mode extend further into the mirror region which increases the optical mode volume, decreasing \gom.

The same technique could be used to create bandgaps at virtually any point in the band diagram by extending the period by a factor of $n$, which would open gaps at $k=m \pi/ n$, $m, n \in \mathbb{N}$. If $n$ is too large however, the effect of tighter mode confinement is negated by the large period. We call this technique \glsxtrlong{xhope} (\glsxtrshort{xhope}).

Since the mode frequency is in the center of the bandgap from the first perturbation, the field starts waning immediately away from the defect. The maximum field strength will thus be in the center (\cref{fig:conv_compare}), where the displacement from the mechanical mode is the largest (\cref{fig:fields}), which further enhances the coupling strength. 
This is qualitatively different from previous implementations \cite{kolvik_clamped_2023}, which gives a wider mode where the intensity peaks at the edges (\cref{fig:conv_compare}), where the field starts to be reflected. This improved spatial overlap between the optical and mechanical field enhances the vacuum optomechanical interaction rate $\gom$. The short mechanical cavity length also results in a clean mechanical spectrum with widely spaced modes, reducing the risk of cluttering the signal with unwanted peaks.


\begin{figure*}[t]
    \centering
    \includegraphics[width=1.0\textwidth]{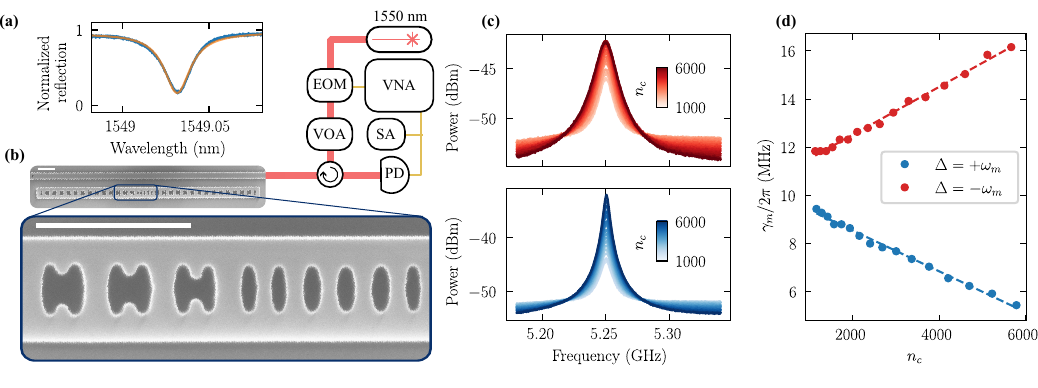}
    \caption{
        \textbf{Measurement of inverse-designed X-HOPE cavity.}
        \textbf{(a)}
        Spectral sweep of optical resonance. Lorentzian fit yields internal quality factor of \num{1.2e5}, with a total linewidth of
        $\kappa / (2\pi) = \qty{2.37}{\giga\hertz}$
        \textbf{(b)}
        Scanning electron micrograph of the device and measurement setup featuring
        a \glsxtrlong{voa}, \glsxtrlong{eom}, \glsxtrlong{pd}, \glsxtrlong{sa}, and \glsxtrlong{vna}.
        \textbf{(c)}
        Mechanical spectrum for different optical powers, quantified as number of intracavity photons~$n_c$.
        \textbf{(d)}
        Fitted mechanical linewidth $\gamma_m$ as function of~$n_c$.
        The slope of the fitted lines give the vacuum optomechanical coupling rates~$\gom/(2\pi) = \qty{770(5)}{\kilo\hertz}$ ($\Delta = +\omega_m$) and $\gom/(2\pi) = \qty{835(10)}{\kilo\hertz}$ ($\Delta = -\omega_m$).
        Scale bars, \qty{1}{\micro\meter}.
    }
    \label{fig:measurement}
    \sublabel{fig:optical_resonance}
    \sublabel{fig:sem_setup}
    \sublabel{fig:mech_spectrum}
    \sublabel{fig:gom}
\end{figure*}

\subsubsection{Inverse design of optomechanical structures}

While the \gls{xhope} technique developed here improves the optomechanical interaction rate, it falls short of delivering a full
optomechanical cavity with desirable optical and mechanical properties.
Specifically, the simulated quality factors are not high enough.
Therefore, we complement \gls{xhope}, which is based on physical reasoning, with a multiphysics inverse-design algorithm to increase performance and bring all properties to acceptable values.

Inverse design has made a large impact on nanophotonic device engineering,
producing devices with smaller footprint and better performance than conventional
design methods \cite{jensen_topology_2011,christiansen_inverse_2021,molesky_inverse_2018,schul_inverse_2026}. 
This class of methods solves the inverse problem: finding a design given desired device properties,
instead of finding the properties of a given design, which is the role of regular simulations.
This can be accomplished in different ways, particle swarm optimization \cite{vitali_highly_2022,kokhanovskiy_inverse_2021}, genetic algorithms \cite{sigmund_topology_2013,jafar-zanjani_adaptive_2018}, and neural networks \cite{jiang_deep_2020,su_metaphynet_2024,deng_neural-adjoint_2021,gahlmann_deep_2022}. In this work, we set up a gradient-based algorithm using adjoint sensitivity analysis \cite{sigmund_topology_2013,molesky_inverse_2018,lalau-keraly_adjoint_2013}, a method which has enabled improved grating couplers \cite{piggott_inverse_2014, michaels_inverse_2018}, frequency demultiplexers \cite{piggott_inverse_2015}, waveguide bends and crossings \cite{shang_inverse-designed_2023}, and beam steering optical phase arrays \cite{vercruysse_inverse-designed_2021} to name a few. While the original method only applies to
problems that reduce to linear systems, e.g.~frequency-domain simulations
\cite{giles_introduction_2000, albrechtsen_nanometer-scale_2022}, an analogous method can be derived for eigenmode simulations \cite{akcelik_adjoint_2005,toader_6_2017}, which are used to simulate the \gls{omc}.

Conventional \glspl{omc} rely on adiabatic transitions between the defect and mirror regions to achieve low scattering loss. This method becomes impractical in situations where smooth transitions are not possible or non-adiabatic transitions are desirable. The abrupt transition that arises between the cells where neighboring holes are separated and where they are merged prevents smooth transitions in our design (\cref{fig:uc_params}). Instead, we optimize the size and position of each hole in the transition region. The result is a non-adiabatic transition that nevertheless maintains a high quality factor. Traditionally, simultaneously optimizing over 200 variables would be prohibitively time consuming, but by implementing an adjoint-based algorithm for computing the gradient of the design parameters with respect to our \gls{fom}, we can use fast, gradient-based optimization methods (\cref{fig:inverse-design}).

We use the single-photon optomechanical scattering rate $\omscat$ as our \gls{fom}.
Additionally, we include constraints on the optical and mechanical mode frequencies, radiation-limited quality factors,
and minimum feature sizes, implemented as penalty terms in the \gls{fom}.
Our \gls{fom} $\fom$ thus depends implicitly on the geometry through both the eigenvalues (frequencies and quality factors)
and the eigenvectors (optical and mechanical fields in expression for $\gom$).
The total (implicit) gradient can be written as a sum of two contributions, one for eigenvalues and one for eigenvectors:
\begin{equation}
    \diff.f.{\fom}{\dvf} = \diff{\fom}{\lambda}\diff.f.{\lambda}{\dvf} + \int \diff.f.{\fom}{\vec v\fop}\diff.f.{\vec v\fop}{\dvf} \iop,
\end{equation}
where $\lambda$ denotes the complex eigenvalue from which the frequency and quality factor is obtained,
and $\vec v$ denotes eigenvector, corresponding to the displacement (electric field) in the acoustic (electromagnetic) simulation, and $\dvf$ denotes the vector of design parameters.
Finally, the gradients for the mechanical and electromagnetic simulations are added together.

To find the gradients of the eigenvalues of an eigenmode simulation, we use
standard methods \cite{fox_rates_1968,lee_adjoint_1999} with discrete adjoints.
To compute gradients for the eigenvector terms, we use a method of continuous adjoints,
similar to references \cite{akcelik_adjoint_2005} and \cite{toader_6_2017} (\cref{fig:inverse-design}).
We run adjoint simulations, which are equivalent to frequency domain simulations with
additional source terms at the eigenfrequency of the respective modes.
The source term is the gradient of the \gls{fom} with respect to the displacement or electric field.
Combining the eigenmode and adjoint fields we obtain the gradient with respect to a material
interpolation field, which can then be translated to a gradient with respect to our design parameters.
Further details are given in the methods section below and in the supplementary text.
Using this gradient, the structure is optimized using the ADAM algorithm \cite{kingma_adam_2017}.
To increase the robustness of the design, the geometry is stochastically eroded or dilated by up
to 2 nm each iteration. This should decrease the probability of becoming trapped in sharp local optima
and increase the tolerance to fabrication imperfections.

The algorithm increases the radiation-limited quality factors at a minor expense of $\gom$
(\cref{fig:optimization_trace}).
As the radiation-limited optical quality factor surpasses \num{1e6} we terminate the
optimization since, by then, the optical quality factor is likely to be limited by surface roughness which is not accounted for in simulation. 

This resulting structure is reminiscent of a topological photonic-crystal nanocavity \cite{ota_topological_2018}. In contrast to \cite{ota_topological_2018}, our design does not connect the two topologically distinct regions directly, but separates them with a defect region where the optomechanical interaction occurs. Moreover, the transition between the mirror and defect is not abrupt  which, in combination with the above inverse design methods, drastically improves quality factors.

The final design has optical and mechanical frequencies of
$\omega_o/(2\pi) = \qty{193.4}{\tera\hertz}$ and $\omega_m/(2\pi) = \qty{5.9}{\giga\hertz}$ respectively.
The optomechanical coupling, taking both photoelastic effect and
radiation pressure into account, is $\gom/(2\pi) = \qty{750}{\kilo\hertz}$,
with radiation pressure being the dominant contribution.
The radiation-limited optical quality factor $Q_o$ is just above \num{1e6} and the radiation-limited mechanical quality factor $Q_m$ is \num{20e3}.

\subsection{Experiment}

We fabricate the optimized \gls{xhope} cavity in \qty{220}{\nm} Si on SiO$_2$ using electron-beam lithography and reactive ion etching. We characterize the device optically and probe the mechanical response with thermal sideband spectroscopy (\cref{fig:sem_setup}) under ambient conditions (see methods section).

The optical resonance has a frequency of $\omega_\text{o}/(2\pi)=\qty{193.5}{\tera\hertz}$ 
and an internal quality factor of $Q_\text{o} = \num{1.2e5}$ (\cref{fig:optical_resonance}). The
mechanical mode is found at $\omega_m/(2\pi) = \qty{5.25}{\giga\hertz}$ with an intrinsic linewidth $\gamma_i/(2\pi)$
of \qty{11}{\mega\hertz} (\cref{fig:mech_spectrum,fig:gom}).
We measure the vacuum optomechanical interaction strength to be $\gom/(2\pi) =
\qty{770(5)}{\kilo\hertz}$ on the blue side ($\Delta = +\omega_m$) and \qty{835(10)}{\kilo\hertz} on the red side ($\Delta= -\omega_m$) (\cref{fig:gom}). 

\begin{figure}
    \centering
    \includegraphics[width=\linewidth]{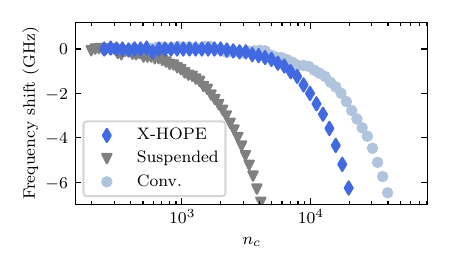}
    \caption{ \textbf{Thermo-optic robustness of the \glsxtrshort{xhope} cavity.} Thermo-optic shift as function of number of intracavity photons $n_c$.
    The \gls{xhope}-design from this paper is compared with the conventionally designed release-free implementation and a suspended \gls{omc} \cite{kolvik_clamped_2023, kolvik_optomechanical_2025}.}
    \label{fig:thermooptics}
\end{figure}

The benefit of the release-free architecture lies in the increased resilience to residual heating at high optical powers.
To quantify this we measure the thermo-optic shift of the optical resonance at different powers.
We find that our device, like previous release-free implementations, remains unaffected at almost 10 times the power where the suspended device begins to redshift (\cref{fig:thermooptics}).

While the experimentally measured values for both the optical frequency and the coupling strength are in good agreement with the
simulations, the measured mechanical frequency is lower than
expected. The simulated radiation-limited quality factors of \num{1e6} and \num{20e3} also exceed the measured values \num{1.2e5} and \num{500} for optical and mechanical modes respectively. We suspect that the optical quality factor is currently limited by scattering associated with the sidewall angle of the structure, while the mechanical quality factor is likely limited by Akhiezer damping in these room-temperature measurements. Indeed, our simulations suggest that the radiation-limited optical quality factor decreases from $10^6$ to $10^5$ with a sidewall angle of \qty{3}{\degree}, which is consistent with the micrograph (\cref{fig:sem_setup}).

\section{Discussion}

In this work, we have demonstrated a release-free optomechanical crystal cavity with a record vacuum optomechanical coupling rate $\gom$, more than 50\% higher than previous release-free implementations. This advance is enabled by the introduction of a cavity design technique for half–X-point optical modes, termed \gls{xhope}, together with a multiphysics inverse-design algorithm for eigenmode simulations. The resulting device achieves optomechanical coupling strengths comparable to those of state-of-the-art suspended optomechanical cavities, while retaining the intrinsic thermal robustness of the release-free architecture.

First-generation release-free \glspl{omc} demonstrated thermal robustness but trailed suspended state-of-the-art devices by roughly a factor of two in vacuum optomechanical coupling \cite{kolvik_clamped_2023, kolvik_optomechanical_2025}, leaving a trade-off between coupling strength and light resilience. In this work, we have largely closed that gap. The release-free silicon \gls{omc} demonstrated here achieves a vacuum optomechanical coupling rate $\gom / (2 \pi) = \qty{800}{\kilo\hertz}$ --- a record for release-free architectures and comparable to state-of-the-art suspended devices --- while retaining their intrinsic thermal robustness. This advance is enabled by combining \gls{xhope}, a cavity design technique for half-X-point optical modes that improves photon--phonon spatial overlap, with a multiphysics inverse-design algorithm that enhances the radiation-limited quality factors.

While the measured coupling rate matches simulation, the measured optical and mechanical quality factors fall short of their radiation-limited values by approximately one and two orders of magnitude, respectively. We attribute these gaps to sidewall scattering and Akhiezer damping at room temperature, neither of which is included in our radiation-limited simulations. Closing these gaps is a near-term priority: cryogenic operation is expected to suppress Akhiezer damping substantially \cite{kolvik_optomechanical_2025}, while improvements in optical quality factor will require fabrication advances or the incorporation of sidewall-angle robustness into the optimization itself. Further increases in the vacuum optomechanical coupling rate $\gom$ may be possible by extending the inverse-design approach to free-form instead of the parameterized optimization employed here, allowing a larger design space to be explored.

Beyond the specific device demonstrated, the release-free \gls{omc} reported here provides the optomechanical building block for the release-free piezo-optomechanical transducer architecture proposed in \cite{burger_design_2025}, and more broadly opens a path toward coherent microwave-to-optical frequency transduction \cite{jiang_optically_2023, mirhosseini_superconducting_2020} and other technologies that leverage gigahertz acoustic waves as an interface between light and platforms such as superconducting qubits. From a wider perspective, both the \gls{xhope} design technique and the multiphysics inverse-design framework introduced here have applicability beyond release-free \glspl{omc}. The \gls{xhope} approach is relevant whenever strong confinement or coupling is required for modes away from the Brillouin-zone edge, including multimode systems with phase-matching constraints such as second-harmonic generation and other three- and four-wave mixing processes. Further, our inverse-design methodology — which handles both eigenvalue and eigenvector sensitivities for a multiphysics figure of merit — applies to any resonant system whose performance depends jointly on mode frequencies, quality factors, and modal overlaps. Concrete targets include co-designed photonic–phononic crystals, piezo-optomechanical and electro-optomechanical resonators, and multimode parametric devices.

\section{Materials and Methods}
\label{sec:methods}

\subsection{Simulation and optimization}

The optomechanical structures are simulated with COMSOL Multiphysics.
The gradients with respect to the eigenvalues of the simulation are obtained using
the built-in sensitivity module.
The gradients of the factors that depend on the fields are obtained with a custom algorithm
which consists of 6 steps (see \cref{fig:id_concept} for a visualization).
\begin{enumerate}[label=(\roman*)]
    \item The geometry is imported into our simulation software. To increase the
        robustness of the final design, the geometry is dilated or eroded by up to 2
        nm \cite{sigmund_manufacturing_2009, wang_robust_2019}. 
    \item An optical and mechanical eigenmode simulation is performed to obtain
        the eigenmodes and (complex) eigenfrequencies.
    \item The gradients of the \gls{fom} w.r.t. the optical and mechanical
        fields are evaluated with an analytically derived formula, see
        the supplementary text for the derivation and full expression.
    \item For both modes, a frequency domain simulation at the eigenfrequency is
        performed where the gradient of the previous step is added as a source
        term. This is the adjoint simulation which gives us the adjoint optical
        and mechanical fields.
    \item The eigenmode fields and the adjoint fields are combined to obtain a
        geometric gradient, indicating where and in which direction the
        \gls{fom} is most sensitive to displacement of the air-silicon
        interface. Again, see the supplementary text for the derivation.
    \item From the geometric gradient, the gradient w.r.t. the design parameters
        of the structure is computed, which is then used to update the parameters
        and obtain a new geometry.
        \label{item:param_grad}
\end{enumerate}

\subsection{Fabrication}

We expose ARP09 2:1 resist with the design pattern using the Raith EBPG 5200 electron
beam lithography system. The developed pattern is etched into a 220 nm silicon
film on 2 µm of silicon oxide with a silicon handle using
a Pseudo-Bosch process with \ce{SF6} as the etching gas and \ce{C4F8} as the passivation gas (Oxford PlasmaPro Deep Reactive Ion Etching).
Subsequently, the samples are cleaned in 3:1 piranha solution and 2~\% HF
before measurement.

\subsection{Measurements}

To probe the device, we couple light from a 1550 nm laser (Santec TSL570) onto
the chip via a grating coupler. The light is guided past the \gls{omc} to a
Bragg grating, where it reflects and travels back to the fiber (\cref{fig:sem_setup}).
The light in the waveguide couples evanescently to modes in the \gls{omc},
and thus we can measure the resonance properties by scanning the laser wavelength over the optical
resonances of the \gls{omc}. We fit a Lorentzian with a polynomial
background to the reflectance dip to compute the resonance frequency and quality
factor (\cref{fig:optical_resonance}).

The mechanical mode is characterized by placing the laser frequency, $\omega_l$, detuned from
the optical resonance frequency, $\omega_o$, by $\Delta \equiv \omega_l - \omega_o = \omega_m$,
the mechanical resonance
frequency. We then create sidebands by modulating the laser with an amplitude modulator
and read out the reflected signal with a high-speed photodetector. A \gls{vna} is
used to sweep the modulation and readout frequencies which lets us fine-tune $\Delta$.
Switching the modulation off, the scattered sideband photons beats
with the pump at the mechanical frequency, and the thermal population of the
mechanical modes is visible in a \gls{sa} connected to the high-speed  (\cref{fig:sem_setup}).

With a blue-detuned pump laser we induce Stokes scattering on the mechanics,
effectively decreasing the linewidth by $4 \gom^2 n_c / \kappa$. A red-detuned pump, in contrast, induces
anti-Stokes scattering which increases the linewidth by the same amount  \cite{aspelmeyer_cavity_2014}. By measuring the
linewidth for different laser powers, we infer the coupling strength \gom (\cref{fig:gom}).

\bibliography{sources}

\section{Acknowledgements}

We thank Joey Frey, Trond Hjerpekj\oe{}n Haug, and Ekaterina Deriushkina for helpful discussions. We acknowledge support from the Knut and Alice Wallenberg foundation through the Wallenberg Centre for Quantum Technology through a WACQT Fellowship, from the European Research Council via Starting Grant 948265, and from the Swedish Foundation for Strategic Research via grant FFL21-0039. The authors declare that they have no competing interests.

\onecolumngrid
\newpage

\counterwithin*{figure}{part}
\stepcounter{part}
\renewcommand{\thefigure}{S\arabic{figure}}
\appendix

\section{Robustness to fabrication imperfections}

\begin{figure}
    \centering
    \includegraphics[]{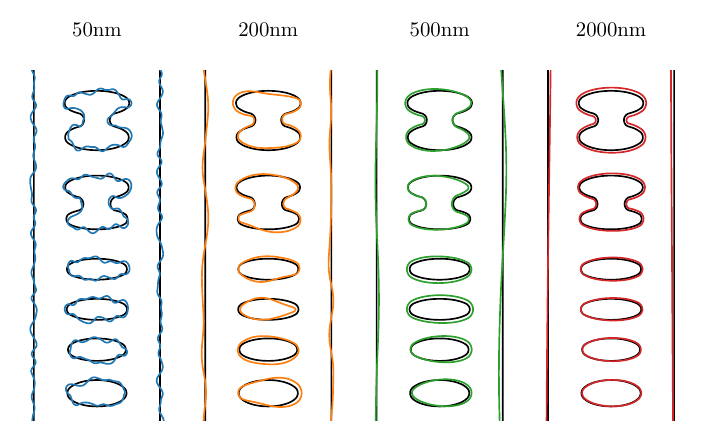}
    \caption{Example perturbations for different length scales with an amplitude of \qty{25}{\nm} for visibility}
    \label{fig:robustness_geom}
\end{figure}
\begin{figure}
    \centering
    \includegraphics[]{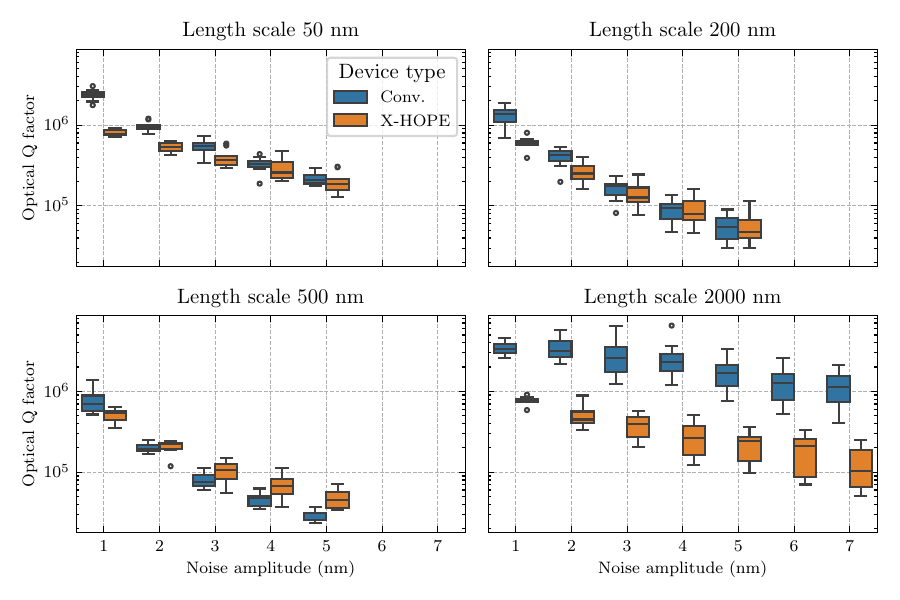}
    \caption{Optical quality factor for different length scales and amplitudes of perturbations. X-HOPE denotes the device design presented in this paper and Conv. the long, first generation release-free \gls{omc} \cite{kolvik_clamped_2023}.}
    \label{fig:robust_box}
\end{figure}
\begin{figure}
    \centering
    \includegraphics[]{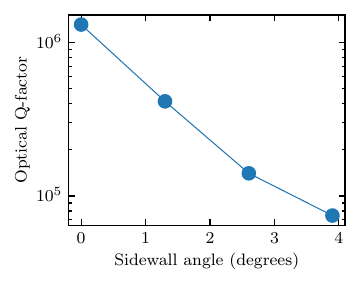}
    \caption{Effect of sidewall angle on optical quality factor in simulation.}
    \label{fig:sidewall}
\end{figure}

We analyze the effect of perturbations to the device geometry in simulation in order to better understand how imperfections in the fabricated devices might impact performance. This is done by adding random offsets to the silicon-air boundary everywhere and performing eigenmode simulations to see how the device is affected. The offsets are generated as OpenSimplex noise over the x-y plane which ensures that the offsets vary smoothly and that both length scale and amplitude of the noise can be controlled (\cref{fig:robustness_geom}).
The offset is constant along the vertical walls of the structure, since that is the form we expect fabrication imperfections to take.

We observe that the radiation-limited optical quality factor of the \gls{xhope} \gls{omc} matches, and in some cases even surpasses the first generation of release-free devices. However, because of the larger initial radiation-limited optical quality factor of the conventional devices, they are still better for perturbation amplitudes smaller than \qtyrange{3}{5}{\nm}.

We also analyze the effect of a non-zero sidewall angle (\cref{fig:sidewall}) and find that the radiation-limited quality factor of our design drops significantly with a few degrees of sidewall, going from \num{1e6} to \num{1e5}. This likely contributes to the lower-than-expected measured optical quality factors.

\section{Design specifics}

Here we describe the specifics of how the design is constructed from the parameters.
The parameters defining our geometry are the width of the beam, the length of the beam,
and the positions and radii of a number of elliptical holes.
The initial design has a transition where all parameters in the transition from mirror to defect follow a cubic polynomial with vanishing initial and final derivative.
In order to remove the bridges between narrow holes we erode the design by \qty{30}{\nm} and then
dilate it again by \qty{30}{\nm}. The first erosion will completely remove bridges with a width less than \qty{60}{\nm},
and the subsequent dilation will not reinstate them since there is no bridge there to dilate.
Features that are larger than \qty{60}{\nm} will be largely unaffected, although sharply curved sections will become less sharply curved.
After this erosion and dilation, the thin bridges will be removed, but the points where they were attached will not be smooth,
so a second round is made where the structure is first dilated by \qty{15}{\nm} and then eroded again by \qty{15}{\nm}.
This enforces a minimum radius of curvature of \qty{15}{\nm}, yielding a smooth design.


\section{Adjoint method derivation and expressions}
\label{app:adjoint}

\subsection{Optomechanical coupling}

We account for two types of optomechanical interactions: one from the moving boundary and the other is from the photoelastic effect.
The moving boundary contribution to the vacuum optomechanical coupling is computed as \cite{safavi-naeini_design_2010}
\[
g_\text{mb} = \frac{\omega_o}{2} \frac{\int_{\partial \Omega} (\vec q \cdot \hat n) (\Delta \epsilon E_{||}^2 - \Delta (\epsilon^{-1}) D_\perp^2) \dl S}{\int \epsilon E^2 \dl V} x_\text{zpf},
\]
where $\omega_o$ is the optical frequency,
$q$ is the normalized displacement of the mechanical mode $q = u / u_\text{max}$,
$\hat n$ is a unit vector along the normal of the surface,
$\Delta \epsilon$ is the difference of permitivities above and below the surface,
$\Delta (\epsilon^{-1})$ is the difference of the inverse of the permitivities above and below the surface,
$E_{||}$ is the electric field parallel to the surface,
$D_{||}$ is the displacement field perpendicular to the surface,
and $x_\text{zpf}$ is the size of the zero-point fluctuations of the mechanical mode.

The photoelastic contribution is computed as
\[
g_\text{pe} = \frac{\omega_o}{2} \frac{\int_{\Omega} \vec E \cdot (\epsilon p S \epsilon) / \epsilon_0 \cdot \vec E \dl V}{\int \epsilon E^2 \dl V}
\]
where $p$ is the rank four photoelastic tensor and S is the rank two strain tensor. 

\subsection{The adjoint method for resonant structures}

Simulation of resonant structures with high quality factors is often done with an eigenvalue solver.
This is fundamentally different from frequency or time domain simulations, which
are both simply solving a linear system of equations $A\vec x = \vec b$. An
eigensolver instead finds solutions $\vec v, \lambda$ to equations like $(M\lambda
^2 + D \lambda + K)\vec v = 0$.

This opens two different variables to optimize: either the eigenvalue $\lambda$,
or the eigenvector $\vec v$.
Since the former has been covered by literature \cite{lee_adjoint_1999, fox_rates_1968}, we omit a repeated description of this here.



\subsubsection{Derivation of the adjoint method for optical eigenmodes}

In the following derivation, we keep the functional form of the fields for as
long as possible, and only discretize when needed. We begin by differentiating
the figure of merit $\fom$ w.r.t. our design variable field $\dvf\fop$.
Computing this is the ultimate goal of this derivation, since this lets us
optimally update $\dvf\fop$.
\begin{equation}
    \diff.f.{\fom(\vec E)}{\dvf\fopp}
    =
    \int \iop \diff.f.{\fom}{\vec E\fop}
    \cdot \diff.f.{\vec E\fop}{\dvf\fopp}
\end{equation}
The first term of the integrand, $\difs.f.{\fom}{\vec E\fop}$, is ``easy'' to
calculate analytically. The second term requires a lot more thought. We first
expand another chain rule.
\begin{equation}
    \diff.f.{\vec E\fop}{\dvf\fopp}
    =
    \int \ioppp \diff.f.{\vec E\fop}{\varepsilon\foppp}
    \diff.f.{\varepsilon\foppp}{\dvf\fopp}
\end{equation}
The design field ultimately does nothing other than determine the dielectric constant everywhere,
and so we can analytically compute $\difs.f.{\varepsilon\fop}{\dvf\fopp}$.
In our case, the material is discretized, so the design parameters shift the silicon-air interface.
The equivalent to increasing $\varepsilon$ by $\Delta \varepsilon =
\varepsilon_\text{Si} - \varepsilon_\text{air}$ in a small volume $\delta V$, i.e. making $\delta V$ into Si,
is to increase $\varepsilon$ by a small amount $\delta\varepsilon$ in an entire mesh element with volume $\Delta V$
if $\Delta \varepsilon \delta V = \delta \varepsilon \Delta V$.

To find $\difs.f.{\vec E\fop}{\varepsilon\fopp}$ we must first consider how the eigenmode equations are solved under the hood.
The equation to be solved is, in the case of electromagnetics,
\begin{equation}
    \nabla \times \frac 1\mu \nabla \times \vec E\fop
    =
    \varepsilon\fop \omega^2 \vec E\fop
\end{equation}
However, solving this using a FEM we usually solve the problem in the so called weak form.
\begin{equation}
    \int \iop \vec t\fop \cdot \nabla \times \frac 1\mu \nabla \times \vec E\fop
    = 
    \int \iop \vec t\fop \cdot \varepsilon\fop \omega^2 \vec E\fop,
\end{equation}
where $\vec t(\vec r)$ is a test function. If this integral equation holds for any test function, then the differential form of the equation must hold.
This can be rewritten in the slightly more convenient form using integration by parts
\begin{equation}
    \label{eq:weak_em}
    \int \iop \nabla \times \vec t\fop \cdot \frac 1\mu \nabla \times \vec E\fop
    - \int_{\partial \Omega} \vec t\fop \times \left(\nabla \times \vec E\fop\right) \cdot \dl{\vec S}
    = 
    \int \iop \vec t\fop \cdot \varepsilon\fop \omega^2 \vec E\fop,
\end{equation}
The surface integral over the outer boundary of the simulation domain can be ignored since the electric field will be 0 there.
Now, we take the functional derivative of this equation w.r.t. $\varepsilon\fopp$:
\begin{equation}
    \int \iop \nabla \times \vec t\fop \cdot \frac 1\mu \nabla \times \diff.f.{\vec E\fop}{\varepsilon\fopp}
    = 
    \int \iop \vec t\fop \cdot \left(\delta(\vec r - \vec r') \omega^2 \vec E\fop
    +
    \varepsilon\fop 2 \omega \diff.f.{\omega}{\varepsilon\fopp} \vec E\fop
    +
    \varepsilon\fop \omega^2 \diff.f.{\vec E\fop}{\varepsilon\fopp}\right).
\end{equation}
Rearranging we get
\begin{equation}
    \label{eq:dif1}
    \int \iop \nabla \times \vec t\fop \cdot \frac 1\mu \nabla \times \diff.f.{\vec E\fop}{\varepsilon\fopp}
    -
    \varepsilon\fop \omega^2 \vec t\fop \cdot \diff.f.{\vec E\fop}{\varepsilon\fopp}
    = 
    \int \iop \vec t\fop \cdot \left(\delta(\vec r - \vec r') \omega^2 \vec E\fop
    +
    \varepsilon\fop 2 \omega \diff.f.{\omega}{\varepsilon\fopp} \vec E\fop
    \right).
\end{equation}
Here comes the ``adjoint'' in the adjoint method: we solve an adjoint problem
\begin{equation}
    \label{eq:adj1}
    \int \iop \nabla \times \tilde{\vec E}\fop \cdot \frac 1\mu \nabla \times \vec t\fop 
    - \varepsilon \fop \omega^2 \tilde{\vec E} \fop \cdot \vec t\fop
    =
    \int \iop \diff.f.{\fom}{\vec E\fop} \cdot \vec t\fop
\end{equation}
You may notice that this is exactly the LHS of \cref{eq:dif1}, but with $\vec t
\mapsto \tilde{\vec E}$ and $\difs.f.{\vec E\fop}{\varepsilon\fopp} \mapsto \vec
t\fop$. We can also see that by just reordering the factors, it is exactly the
same as \cref{eq:weak_em}, although this time, the frequency is known. From this
we thus see that solving this adjoint equation is nothing more than solving a
normal frequency domain simulation, albeit with a complex frequency. The
imaginary part of the complex frequency is equivalent to adding loss to the
structure everywhere. At a glance, the equation may seem ill-defined, since the
operator $\varepsilon\omega^2 - \nabla\times\mu^{-1}\nabla\times$ is, by the
definition of $\omega$, non-injective. Thus the equation cannot be solved for an
arbitrary RHS. However, since $\fom$ is independent of the scale of $\vec
E\fop$, e.g. $\fom(\vec E\fop) = \fom(2\vec E\fop)$ which must be the case since
the magnitude of the eigenmode field is arbitrary, $\difs.f.{\fom}{\vec E\fop}$
must be orthogonal to $\vec E\fop$. This ensures that the equation indeed is
solvable.

Since \cref{eq:adj1} holds for any test function $\vec t\fop$,
it holds for $\vec t\fop = \difs.f.{\vec E\fop}{\varepsilon\fopp}$.
Replacing the LHS of \cref{eq:dif1},
where $\vec t\fop$ is replaced by $\tilde{\vec E}\fop$,
with the RHS of \cref{eq:adj1},
where $\vec t\fop$ is replaced by
$\difs.f.{\vec E\fop}{\varepsilon\fopp}$,
we get
\begin{equation}
    \int \iop \diff.f.{\fom}{\vec E \fop} \cdot \diff.f.{\vec E \fop}{\varepsilon\fopp}
    =
    \int \iop \tilde{\vec E}\fop \cdot \left(
    \delta(\vec r - \vec r') \omega^2 \vec E\fop
    +
    \varepsilon\fop 2 \omega \diff.f.{\omega}{\varepsilon\fopp} \vec E\fop
    \right).
\end{equation}
Thus we find that
\begin{equation}
    \diff.f.{\fom}{\varepsilon\fopp} 
    =
    \omega^2 \tilde{\vec E}\fopp \cdot \vec E\fopp
    + 2 \omega \diff.f.{\omega}{\varepsilon\fopp} 
    \int \iop \varepsilon\fop \tilde{\vec E}\fop \cdot \vec E\fop.
\end{equation}

One final note is that 
$\fom$ often isn't a holomorphic function,
often depending on both $\vec E$ and it's complex conjugate $\vec E^*$,
for example $\abs{\vec E}^2 = \vec E \cdot \vec E^*$.
In such a case, differentiating w.r.t. the complex argument is not well defined.
A simple workaround is to treat $\vec E$ and $\vec E^*$ as separate independent variables, and thus compute
\begin{equation}
    \diff{\fom}{\varepsilon\fopp}
    =
    \int \iop
    \diff.f.{\fom}{\vec E\fop} \cdot \diff.f.{\vec E\fop}{\varepsilon\fopp}
    + \diff.f.{\fom}{\vec E^*\fop} \cdot \diff.f.{\vec E^*\fop}{\varepsilon\fopp}
\end{equation}
It is obvious that $\difs.f.{\vec E^*\fop}{\varepsilon\fopp} = (\difs.f.{\vec E\fop}{\varepsilon\fopp})^*$,
and, though it is less obvious, it is also true that $\difs.f.{\fom}{\vec E^*\fop} = (\difs.f.{\fom}{\vec E\fop})^*$
if $\fom$ is real-valued.
If the above holds, the second term is simply the conjugate of the first and the sum is twice the real part of the first term.
\begin{equation}
    \diff{\fom}{\varepsilon\fopp}
    =
    2 \Re \left\{ \int \iop
    \diff.f.{\fom}{\vec E\fop} \cdot \diff.f.{\vec E\fop}{\varepsilon\fopp}
    \right\}
\end{equation}
\subsubsection{Derivation of the adjoint method for acoustic eigenmodes}

The governing equation for acoustic eigenmode simulations is a little bit different,
but the parallels to electromagnetics are clear: this is also a wave equation where the density $\rho$ corresponds to
electric permittivity and the stiffness tensor $C_{ijkl}$ corresponds the magnetic permeability.
Since the equations now get more dimensions (the stiffness tensor is rank four) we switch to using index notation.
\begin{equation}
    \partial_j \left[ C_{ijkl}\fop \partial_l u_k\fop\right] + \rho\fop \omega^2 u_i\fop = 0
\end{equation}
Just like before, the equation is solved in the weak form:
\begin{equation}
    \int \iop t_i\fop \partial_j C_{ijkl}\fop \partial_l u_k\fop + \rho\fop \omega^2 t_i\fop u_i\fop
    = 0
\end{equation}
which we again rewrite using integration by parts:
\begin{equation}
    \int \iop t_i\fop \partial_j C_{ijkl}\fop \partial_l u_k\fop
    =
    \int \iop \left(\partial_j t_i\right) C_{ijkl}\fop \partial_l u_k\fop
    +
    \int_{\partial \Omega} t_i\fop C_{ijkl} \partial_l u_k\fop n_j \dl S
\end{equation}
In the final integral, $n_j$ is the unit vector normal to the surface element.
The boundary is more complex in the solid mechanics simulations than the the electromagnetic,
but because we are using traction-free boundary conditions, i.e. $n_jC_{ijkl}\partial_l u_k = \sigma_{ij} n_j = 0$, the final term is 0.
Thus we get
\begin{equation}
    \label{eq:weak_mech}
    \int \iop \left(\partial_j t_i\right) C_{ijkl}\fop \partial_l u_k\fop
    + \rho\fop \omega^2 t_i\fop u_i\fop
    = 0
\end{equation}

For mechanics, both $C_{ijkl}\fop$ and $\rho\fop$ depend on the design variable field $\dvf$.
Thus, the chain rule gives
\begin{equation}
    \diff.f.{\fom}{\dvf\fop} = \int \iopp \int \ioppp
    \diff.f.{\fom}{u_i\fopp} \diff.f.{u_i\fopp}{\rho\foppp} \diff.f.{\rho\foppp}{\dvf\fop}
    +\diff.f.{\fom}{u_i\fopp} \diff.f.{u_i\fopp}{C_{jklm}\foppp} \diff.f.{C_{jklm}\foppp}{\dvf\fop}
\end{equation}
Let's start with $\rho$.

Differentiating \cref{eq:weak_mech} w.r.t. $\rho\fopp$, we obtain
\begin{equation}
\begin{split}
    \label{eq:dif2}
    \int \iop \left(\partial_j t_i\fop\right) C_{ijkl}\fop \partial_l \diff.f.{u_k\fop}{\rho\fopp}
    +& \rho\fop \omega^2 t_i\fop \diff.f.{u_i\fop}{\rho\fopp}\\
    =& - \int \iop \delta(\vec r - \vec r') \omega^2 t_i \fop u_i\fop + \rho\fop 2 \omega \diff.f.{\omega}{\rho\fopp} t_i\fop u_i\fop\\
    =& - \omega^2 t_i \fopp u_i\fopp - 2 \omega \diff.f.{\omega}{\rho\fopp} \int \iop \rho\fop  t_i\fop u_i\fop
\end{split}
\end{equation}
Then we again solve for an adjoint field satisfying the equation
\begin{equation}
    \label{eq:adj2}
    \int \iop \left(\partial_j \tilde u_i\fop\right) C_{ijkl}\fop \partial_l t_k\fop
    + \rho\fop \omega^2 \tilde u_i\fop t_i\fop
    = \int \iop t_i\fop \diff.f.{\fom}{u_i\fop}
\end{equation}
Noting that $C_{ijkl} = C_{klij}$ we see that this is the same as a frequency domain simulation
with the RHS as the source term.
Now we do the same trick as before:
identify $t_i$ in \cref{eq:dif2} with $\tilde u_i$ in \cref{eq:adj2}
and $t_k\fop$ in \cref{eq:adj2} with $\difs.f.{u_k\fop}{\rho\fopp}$.
Thus we get
\begin{equation}
    \int \iop \diff.f.{\fom}{u_i\fop} \diff.f.{u_i\fop}{\rho\fopp}
    =
    - \omega^2 \tilde u_i \fopp u_i\fopp
    - 2 \omega \diff.f.{\omega}{\rho\fopp} \int \iop \rho\fop  \tilde u_i\fop u_i\fop
\end{equation}

Doing the same for $C_{ijkl}$ changes \cref{eq:dif2} to
\begin{equation}
\begin{split}
    \int \iop \left(\partial_j t_i\fop\right) & C_{ijkl}\fop 
    \partial_l \diff.f.{u_k\fop}{C_{i'j'k'l'}\fopp}
    + \rho\fop \omega^2 t_i\fop \diff.f.{u_i\fop}{C_{i'j'k'l'}\fopp}\\
    =& - \int \iop \left(\partial_j t_i\fop\right) 
    \delta(\vec r - \vec r') \delta^{ijkl}_{i'j'k'l'}\partial_l u_k \fop
    - 2 \omega \diff.f.{\omega}{C_{i'j'k'l'}\fopp} \rho\fop  t_i\fop u_i\fop\\
    =& - \left(\partial_{j'} t_{i'}\fopp\right) \partial_{l'} u_{k'} \fopp 
    - 2 \omega \diff.f.{\omega}{C_{i'j'k'l'}\fopp} \int \iop \rho\fop  t_i\fop u_i\fop
\end{split}
\end{equation}
Thus we get
\begin{equation}
    \int \iop \diff.f.{\fom}{u_m\fop} \diff.f.{u_m\fop}{C_{ijkl}\fopp}
    =
    - \left(\partial_j \tilde u_i\fopp\right) \partial_l u_k\fopp
    - 2 \omega \diff.f.{\omega}{C_{ijkl}\fopp} \int \iop \rho\fop  \tilde{u}_i\fop u_i\fop
\end{equation}

\subsubsection{Computing the source terms}

The above gives the general method for computing the gradient.
The only specific derivation needed for a given figure of merit $\fom$
is to get an expression for $\difs.f.{\fom}{\vec E\fop}$
and $\difs.f.{\fom}{u_i\fop}$ to us as a source term in the adjoint simulations (equations \cref{eq:adj1,eq:adj2}).
Below we derive these terms for $\fom \equiv g_\text{mb}$.
obtaining them for other simple expressions with $g_\text{mb}$,
such as $\abs{g_\text{mb}}^2$ will then be trivial.

The electric field appears 3 times in the expression for the moving boundary coupling.
The functional derivative of the parallel component is
\begin{align}
    \diff.f.{E_\parallel^2\fop}{\vec E\fopp} &= 
    \vec E_\parallel^* \cdot \diff.f.*{
        \left(\vec E\fop - (\hat{\vec n} \cdot \vec E\fop) \hat{\vec n}\right)
    }{\vec E\fopp}\\
    &=
    \vec E_\parallel^* \cdot
    \delta(\vec r - \vec r') (I  - \hat{\vec n}\hat{\vec n} )\\
    &= 
    \vec E_\parallel^* \cdot
    \delta(\vec r - \vec r').
\end{align}
Similarily,
\begin{equation}
    \diff.f.{D_\perp^2\fop}{\vec E\fopp} = 
    \epsilon \fop \vec D_\perp^*
    \delta(\vec r - \vec r').
\end{equation}
and of course
\begin{equation}
    \diff.f.{E^2\fop}{\vec E\fopp} = 
    \vec E^*\fop
    \delta(\vec r - \vec r').
\end{equation}

Thus we get
\begin{align}
    \diff.f.{g_\text{mb}}{\vec E\fopp}
    =& 
    x_\text{zpf} \frac{\omega_o}{2}
    \frac{\int_{\partial \Omega} 
        \delta(\vec r-\vec r')
        (\vec q \cdot \hat n)
        (\Delta \epsilon \vec E_{||}^* - \Delta \epsilon^{-1} \vec D_\perp^*) \dl S
    }{\int \epsilon E^2 \dl V} \\
    &- 
    x_\text{zpf} \frac{\omega_o}{2} 
    \frac{\int_{\partial \Omega}
        (\vec q \cdot \hat n) 
        (\Delta \epsilon E_{||}^2 - \Delta \epsilon^{-1} D_\perp^2) 
        \dl S
    }{\left(\int \epsilon E^2 \dl V\right)^2}
    \int \epsilon \vec E^* \fop \delta(\vec r - \vec r') \dl V\\
    =& 
    x_\text{zpf} \frac{\omega_o}{2}
    \frac{\int_{\partial \Omega} 
        \delta(\vec r-\vec r')
        (\vec q \cdot \hat n)
        (\Delta \epsilon \vec E_{||}^* - \Delta \epsilon^{-1} \vec D_\perp^*) \dl S
    }{\int \epsilon E^2 \dl V} \\
    &- 
    x_\text{zpf} \frac{\omega_o}{2} 
    \frac{\int_{\partial \Omega}
        (\vec q \cdot \hat n) 
        (\Delta \epsilon E_{||}^2 - \Delta \epsilon^{-1} D_\perp^2) 
        \dl S
    }{\left(\int \epsilon E^2 \dl V\right)^2}
    \epsilon\fopp \vec E^* \fopp \dl V
    \label{eq:dgde}
\end{align}
Thus we get two terms in the RHS of \cref{eq:adj1}:
one surface term, the $\delta$ in the first term effectively replaces the integral in \cref{eq:adj1} by a surface integral, and one volume term.
Using the figure of merit $\fom = \abs{g_\text{mb}}^2 = g_\text{mb}^* g_\text{mb}$ gives
\begin{equation}
    \diff.f.{\abs{g_\text{mb}}^2}{\vec E\fop} = 
    \diff.f.{g_\text{mb}}{\vec E\fop} g_\text{mb}^*
    + \diff.f.{g_\text{mb}^*}{\vec E\fop} g_\text{mb}
\end{equation}
It can be shown that 
$\difs.f.{g_\text{mb}}{\vec E\fop} = \difs.f.{g_\text{mb}^*}{\vec E\fop} $ if one replaces $\vec q$ with $\vec q^*$ in \cref{eq:dgde}.

The corresponding term for the mechanical mode is computed in a similar way. Remember that 
$\vec q = \vec u / u_{\max} $, and
$x_\text{zpf} = \sqrt{\hbar / 2 m_\text{eff} \omega_m}$
where 
$m_\text{eff} = \int \rho\fop q^2\fop \iop$.
Also note that since we normalize by the zero-point fluctuations, $g_\text{mb}$ is independent of multiplying $\vec q$ by a scalar.
Thus, differentiating w.r.t. $\vec q$ and $\vec u$ is the same thing except a scale factor of $1/u_{\max}$.
\begin{align}
    u_{\max}\diff.f.{g_\text{mb}}{\vec u\fopp} =& 
    \diff.f.{g_\text{mb}}{\vec q\fopp} \\
    =&
    -\sqrt{\frac{\hbar}{2 \omega_m}}
    \frac12 m_\text{eff}^{-3/2}
    \int \rho\fop \vec q^*\fop \delta(\vec r - \vec r') \iop
    \cdot \frac{\omega_o}{2} 
    \frac{\int_{\partial \Omega}
        (\vec q \cdot \hat n) 
        (\Delta \epsilon E_{||}^2 - \Delta \epsilon^{-1} D_\perp^2) 
        \dl S
    }{\int \epsilon E^2 \dl V}\\
    &+ 
    x_\text{zpf} \frac{\omega_o}{2} 
    \frac{\int_{\partial \Omega}
        (\delta(\vec r - \vec r') \hat n) 
        (\Delta \epsilon E_{||}^2 - \Delta \epsilon^{-1} D_\perp^2) 
        \dl S
    }{\int \epsilon E^2 \dl V}\\
    =&
    -\sqrt{\frac{\hbar}{2 \omega_m}}
    \frac12 m_\text{eff}^{-3/2}
    \rho\fopp \vec q^*\fopp
    \cdot \frac{\omega_o}{2} 
    \frac{\int_{\partial \Omega}
        (\vec q \cdot \hat n) 
        (\Delta \epsilon E_{||}^2 - \Delta \epsilon^{-1} D_\perp^2) 
        \dl S
    }{\int \epsilon E^2 \dl V}\\
    &+ 
    x_\text{zpf} \frac{\omega_o}{2} 
    \frac{\int_{\partial \Omega}
        (\delta(\vec r - \vec r') \hat n) 
        (\Delta \epsilon E_{||}^2 - \Delta \epsilon^{-1} D_\perp^2) 
        \dl S
    }{\int \epsilon E^2 \dl V}
\end{align}


\end{document}